\documentclass[a4paper,UKenglish,cleveref, autoref, thm-restate]{oasics-v2021}

\nolinenumbers
\usepackage[T1]{fontenc}

\usepackage[utf8]{inputenc}
\usepackage[cachedir=.,finalizecache=false ,frozencache=true]{minted}

\usepackage[pdftex]{xcolor}
\usepackage{todonotes}

\usepackage{amsmath}
\usepackage{amsfonts}
\usepackage{amssymb}
\usepackage{mathtools}

\usepackage{paralist}

\usepackage[ligature,reserved]{semantic}

\mathlig{++}{\cdot}
\reservestyle{\listT}{\texttt}
\listT{cons, nil}

\newcommand\solin[1]{\mintinline{solidity}{#1}}
\newcommand\coqin[1]{\mintinline{coq}{#1}}

\newcommand\sto{\ensuremath{\to}}
\newcommand\sand{\ensuremath{\land}}
\newcommand\sequiv{\ensuremath{\equiv}}

\newcommand{\myparagraph}[1]{\noindent\textsf{\textbf{#1}}}

\title{Multi: a Formal Playground for \\Multi-Smart Contract
  Interaction}

\author{Martín Ceresa}{IMDEA Software Institute, Madrid}{martin.ceresa@imdea.org}{https://orcid.org/ 0000-0003-4691-5831}{}%
\author{César Sánchez}{IMDEA Software Institute, Madrid}{cesar.sanchez@imdea.org}{https://orcid.org/ 0000-0003-3927-4773}{}

\authorrunning{Martín Ceresa and César Sánchez}
\titlerunning{Multi: a Formal Playground for Multi-Smart Contract Interaction}
\Copyright{Martín Ceresa and César Sánchez}

\funding{This work was funded in part by the Madrid Regional Government
  under project ``S2018/TCS-4339 (BLOQUES-CM)‘’ and by a research
  grant from Nomadic Labs and the Tezos Foundation.}

\relatedversiondetails[cite=Ceresa.2022.Multi]{Full Version}{} 

\keywords{blockchain, formal methods, theorem prover, smart-contracts}

\ccsdesc[500]{Theory of computation~Program reasoning}
\ccsdesc[500]{Theory of computation~Program constructs}
\ccsdesc[500]{Theory of computation~Abstract machines}

\begin{document}

\maketitle

\begin{abstract}
%
%
Blockchains are maintained by a network of participants, miner nodes, that run
algorithms designed to maintain collectively a distributed machine tolerant to
Byzantine attacks.
From the point of view of users, blockchains provide the illusion of centralized
computers that perform trustable verifiable computations, where all computations
are deterministic and the results cannot be manipulated or undone.

Every blockchain is equipped with a crypto-currency.
Programs running on blockchains are called smart-contracts and are
written in a special-purpose programming language with deterministic
semantics\footnote{Although the behaviour of smart-contracts may depend on
values to be known at runtime, i.e. block number; hashes; etc, their behaviour
is deterministic.}.
Each transaction begins with an invocation from an external user to a smart
contract.
%
Smart contracts have local storage and can call other contracts, and more
importantly, they store, send and receive cryptocurrency.

Once installed in a blockchain, the code of the smart-contract cannot
be modified.
Therefore, it is very important to guarantee that contracts are
correct before deployment.
However, the resulting ecosystem makes it very difficult to reason
about program correctness, since smart-contracts can be executed by
malicious users or malicious smart-contracts can be designed to
exploit other contracts that call them.
Many attacks and bugs are caused by unexpected interactions between multiple
contracts, the attacked contract and unknown code that performs the exploit.

Moreover, there is a very aggressive competition between
different blockchains to expand their user base.
Ideas are implemented fast and blockchains compete to offer and adopt
new features quickly.

In this paper, we propose a \emph{formal playground} that allows
reasoning about multi-contract interactions and is extensible to
incorporate new features, study their behaviour and ultimately prove
properties before features are incorporated into the real blockchain.
We implemented a model of computation that models the execution
platform, abstracts the internal code of each individual contract and
focuses on contract interactions.
Even though our Coq implementation is still a work in progress, we show
how many features, existing or proposed, can be used to reason about
multi-contract interactions.
\end{abstract}


\section{Introduction}

%
%
Smart-contract manipulate cryptocurrency, which has a corresponding
value as money.
Since smart-contracts cannot be modified once installed and their
computations cannot be undone (``the contract is the law''), all
interactions with the contract are considered valid.
Therefore, there is an incentive for malicious users to take advantage
from unexpected behaviors and interactions.
Also, errors in contracts can result in losses and cryptocurrency
being locked indefinitely, even when used but by well-intentioned
users.
We focus in this paper on the computational notion of correctness, and
not on the real legal implications resulting from interactions in the
blockchain or the use of smart-contracts to enforce legally binding
contracts~\cite{Ellul.2020.RegulatingBlockchain}.

%
%
One important reason why it is very difficult to reason about smart
contracts is that they live in an \emph{open universe}.
Even though the code of a given smart-contract $C$ cannot be modified
once installed, other contracts that call and are called from $C$ can
be programmed and deployed after malicious users study $C$.
Therefore, programmers and auditors of contract $C$ did not have to
analyze all possible code that can invoke or be invoked from
$C$.

%
%
At the same time, users demand blockchains to implement new
features.
Since there is a big competition between blockchains, this
puts pressure on architects of blockchains on the time to market
of new features.
And each new feature potentially increases the attack surface of smart
contracts.

There are different kinds of errors found in smart-contracts.
\begin{compactitem}
\item \emph{Logical problems} are related to errors in the logic of
  the smart-contract.
Usually, attackers detect a corner case that can be exploited to
generate an unwanted behaviour.
\item \emph{Low-level execution} problems that arise from a misunderstanding
on details of the low-level execution platform.
Examples include underflow, overflow or exploiting unexpected behavior
after the stack limit is reached.
\item Programmer can also employ \emph{bad idioms} that they are
  familiar with from other areas of software applications, but which
  may be dangerous in interactive platforms like blockchains, where
  all data (including the state of the contracts) is
  public and verifiable.
\end{compactitem}
Most bugs are related to multi-contract interactions.
For example, the infamous DAO attack where malicious code \emph{legally}
exploited the machinery of the Ethereum blockchain creating unexpected
re-entrant calls from remote contracts led to the loss of
\$$60$ million~\cite{Daian.2016.DAO}.

%
%
In this article, we present a formalization in Coq of a general
blockchain model of computation that allows us to study new
multi-contract interactions as well as new features.
We aim to develop a formal and rigorous way to analyze the possible
interactions between contracts and also to study how new features
affect contracts before they are implemented and deployed.
Our Coq library allows simulating the execution of smart-contracts,
abstracting away the internal code of the contract.
Our abstraction is based on the Tezos blockchain, but it is general
enough to cover other blockchains like Ethereum.
We model smart-contract (almost) as pure functions from the current
storage and state of the blockchain into (possibly) a list of
operations to do next plus changes in the storage.


\section{Motivation}

After successful attacks like DAO~\cite{Daian.2016.DAO} there is a growing
interest in formal methods for smart-contracts.
First, there is an interest in verifying that a contract satisfies a
specification so certain properties can be guaranteed, e.g.  the owner will be
able to fetch all funds or that a bidder will either gain the bidding or recover
the funds.
Second, it is also important to formally study different mechanisms and features
proposed for a given blockchain before they are offered so new attacks can be
prevented.
Some of these mechanisms are proposed to allow users to use more
effective defensive programming idioms.

For example, by analyzing the DAO
attack~\cite{Grossman.2018.EffectivelyCallbackFree} proposed a
property called \emph{effectively callback free} which restricts the
interactions within smart-contracts disabling these attacks.
Later on, the Tezos blockchain~\cite{Allombert.2019.Tezos} implements such
property by construction: smart-contracts are functions that either fail or
returning a list of operations to be executed plus a new storage.
Therefore, the storage is updated before the operations are executed,
which prevents attacks like the DAO using this programming style.

In order to prevent these attacks, the Tezos blockchain followed a
conservative scheduling strategy.
In Tezos, as is the general case, every transaction begins with a request by an
external user indicating the smart-contract to invoke, method and arguments, and
balance of the initial operation.
Assume user \(Alice\) starts a transaction invoking method \(f\)
of smart-contract \(C\), and that, after executing \(C.f\) we have a
list of operations \([o_{0}, \ldots, o_{n}]\).
To compute the result of the transaction, the blockchain will execute
each operation \(o_{i}\) in order, until the gas is exhausted or the
list of pending operations is empty.
%
%
The order in which the operations are executed affects the outcome
of the transaction.
Two conventional strategies are: (1) to insert the new list of
operations at the beginning of the list of pending operations (DFS)
(2) to insert the new list of operations at the end (BFS).
The first one, DFS, allows us to implement a call-and-return flow of
computation and it is the more conventional in most blockchains.
The second one, BFS, prevents call injection attacks by construction as one can
guarantee that two operations are executed back-to-back and was used
until version~\(8\)~ of Tezos~(Protocol Edo)~\cite{Tezos.Full.DFS}.
%
%
In our example, assuming that executing \(o_{1}\) generates \(\mathit{bs}\)
operations, the result of the previous execution would be
\([o_{2}, \ldots, o_{n}] ++ bs\).
While in DFS, the result would be
\( \mathit{bs} ++ [o_{2}, \ldots, o_{n}]\), and thus, the instructions in
\(\mathit{bs}\) will be executed before \(o_{2}, \ldots, o_{n}\).
%
However, BFS suffers from other classes of problems.

%
%
Assume a bank contract that holds money for a customer and the bank contract is
willing to send money as long as the balance stays above
threshold~\mintinline[]{solidity}/threshold/.
In a solidity like language, the contract could be as follows:

\begin{minted}{solidity}
contract Bank {
  uint threshold;
  address owner;
  constructor(uint _threshold, address _owner) public {
    threshold = _threshold;
    owner = _owner;
  }
  function deposit() payable public{
    return([]);
  }
  function withdraw(uint ret) public {
    if (sender = owner) then
        if (balance - ret > threshold) then
            return ([transfer(owner.Receive, ret)])
        else
            fail("breaking invariant")
    else
        fail("not owner")
  }
}
\end{minted}
Normal usage of a such a bank contract can be:
\begin{minted}{solidity}
contract GoodClient{
    address bank;
    // ...
    function askMoney(uint m){  // Requests m from the vault
        return([bank.withdraw(m)]);
    }
}
\end{minted}
%
%
On the other hand, the following is a simple attack exploiting the bank
contract:
\begin{minted}{solidity}
contract Bad{
   address bank;
   //...
   function rob(uint n, uint m){ // BFS attack to the vault!
       return(ntimes n [bank.withdraw(m)])
   }
}
\end{minted}
%
The new method called \mintinline{solidity}{rob} generates a
list of invocations to the vault.
Assume the vault contract has a threshold of \mintinline{solidity}{9} and that
is in a state in which it stores \solin{15} units of cryptocurrency.
A simple examination suggests that the vault will send money back to its owner
whenever its balance is greater than \solin{9}, effectively allowing only one
withdrawal.
However, consider the following execution starting from
$\mintinline{solidity}{[rob(3,5))]}$.
After executing the operations, we would have the following pending queue:
\[ \small
  \solin{[(Bad, vault.withdraw(5)),}
  \solin{ (Bad, vault.withdraw(5)),} 
  \solin{ (Bad, vault.withdraw(5))]} 
\]
Then the BFS sequence of executions leads to the following sequence of
pending operations:
\[ \small \arraycolsep=0.4em
  \begin{array}{llll}
  \solin{[(Bad, vault.withdraw(5)),} &
  \solin{(Bad, vault.withdraw(5)),} &
  \solin{(Bad, vault.withdraw(5))]} & \leadsto \\
  \solin{[(Bad, vault.withdraw(5)),} &
  \solin{(Bad, vault.withdraw(5)),} &
  \solin{(Vault, Bad.Receive())]}  & \leadsto \\
  \solin{[(Bad, vault.withdraw(5)),} &
  \solin{(Vault, Bad.Receive()),} &
  \solin{(Vault, Bad.Receive())]} & \leadsto \\
  \solin{[(Vault, Bad.Receive()),} &
  \solin{(Vault, Bad.Receive()),} &
  \solin{(Vault, Bad.Receive())]} & \leadsto \\
  \solin{[(Vault, Bad.Receive()),} &
  \solin{(Vault, Bad.Receive())]} && \leadsto \\
  \solin{[(Vault, Bad.Receive()) ]} &&& \leadsto \\
  \solin{[]}
\end{array}
\]

First, the operation sending the money back to contract \solin{Bad}
is added at the end, as dictated by BFS.
Second, according to the semantics of feature ``transfer'' in the
Tezos blockchain, funds are subtracted from the sending contract
\solin{Vault} after the transfer is executed.
Therefore, the second \solin{withdraw} request does not see the effect
of attending the first one.
%
%
%
The combined effect is that all three requests are attended resulting
in a total extraction of $15$ units leaving $0$ in contract \solin{Vault}
\emph{without noticing the attack}.
The attack is based on the separation between the creation of a
transfer and its execution.
The lesson is that even though a BFS order prevents injection
attacks, it allows attacks based on the delayed effect of emitted
operation.
The contract \solin{Vault} can be easily fixed by encoding in a
variable in the storage the balance that has been compromised with a
future transfer.
If necessary, \solin{withdraw} can create two operations (1) the
transfer, and (2) an invocation to a new private method in
\solin{Vault} whose purpose is to note that the compromised
balance created by a withdraw has been effectively arrived.


Another lesson is that relying on the balance of contracts is
considered a bad smart-contract programming practice.
Assume now that programmers would like the architects of the
blockchain to implement not only \solin{balance} but also
\solin{pending_balance}, which accounts for transfers sent but not
executed.
Moreover, assume also that the blockchain also implements the feature of
\emph{views}, an apparently innocent feature that simply returns information
about the storage of a contract without any effect.
We illustrate that these two features combined can lead to undesirable
effects.
For example, if we would like to maintain the invariant that at every
moment the amount of combined funds between a collection of contracts
is constant, the combination of \solin{pending_balance} and views can
break such an invariant.

For example, consider three smart-contract \(A,B,C\), and the
following pending queue of operations:
\[
  [ \underbrace{A_{1}, \ldots, A_{o}}_{A} , \underbrace{C_{1}, \ldots, C_{m}}_{C} , \underbrace{B_{1} ,\ldots , B_{n}}_{B}]
\]
where \(A\) sends money to \(B\)---in operations that are going to be
executed after \(C\) but that update \(A\) pending balance.
This leaves $C$ in a difficult position.
If \(C\) observes (using views) the balances of \(A\) and \(B\) there
is going to be a mismatch with their real balances, because $C$ will
see the pending compromised balance but not the pending receives,
which may induce bad behaviour in \(C\).
If \(C\) depends on \(A.balance + B.balance\), for example, to buy some NFT it
may incorrectly fail to take the right decision.
A possible solution is to introduce yet another feature that captures
pending receives.

In our line of work, we aim to build a \emph{formal playground} where different
features and mechanisms can be encoded and reasoned about easily and formally,
while also simulating the execution of multiple contracts.


\section{Previous Work}

In our work, we follow a static verification approach where contracts and
features are analyzed before deployment.
The idea is to encode how blockchains are implemented and study the
behavior of contracts and features by formally proving properties.
Several approaches have been suggested for testing, model checking
and functional and temporal verification of smart-contracts.
We review the most relevant.

\vspace{0.5em}
\myparagraph{Mi-Cho-Coq.}
Mi-Cho-Coq is the first verification tool implemented in Coq for the
Tezos blockchain ecosystem~\cite{Bruno.2019.MiChoCoq}.
The main difference between Mi-Cho-Coq and our effort (Multi) is that
Mi-Cho-Coq focuses on the analysis of the code of a single
contract (or collection of calling contracts for which the code is
available).
We say that Mi-Cho-Coq implements small-step semantics to prove
\emph{functional properties}, which requires to have a concrete
specification of a smart-contract  and either its code or a higher
level specification.

The main difference with Mi-Cho-Coq is that our goal is to prove
properties \emph{emerging} from interactions between smart-contracts.
Our tool is a complementary effort to lower-level verification tools
as Mi-Cho-Coq.

\vspace{0.5em}
\myparagraph{Concert.}
Concert~\cite{Annenkov.2020.ConCert} is another framework written in
Coq to prove formal properties of smart-contracts, and in this case,
they accept multi-contract
interaction~\cite{Nielsen.2019.SCInteractions}.
The fundamental idea of Concert is to model of smart-contracts as
agents and computation as interaction (message passing) between these
agents.
They also implement specific mechanisms, for example,
they implement delegation primitives in the Tezos blockchain.
%
Moreover, Concert has an extraction mechanism to extract high-level
smart-contracts written in Ligo~\cite{LigoLang}.
%

Our main difference is that we implement a very flexible framework with the idea
of encoding new potential blockchain features and prove properties of how
different features interact with each other.
Including BFS and DFS scheduling in the Tezos blockchain, but
there may be other scheduling strategies.

Concert implements blockchains in a generic way using specific
features of Coq (class system) and meta-programming features to easily
embed blockchain smart-contract languages.
Concert also builds proofs by inspecting the trace representing the
evolution of the blockchain observed by a \emph{small step relation}.

Implementing new blockchain features relating to how smart-contracts are
executed is an important feature in our framework, and moreover, we want to be
able to reason and prove properties about such features.
For example, what would happen if smart-contracts can inspect
\emph{runtime} information as the stack call (what the next operations
or pending operations are).
Another difference is that (so far) we observe the state of the
blockchain comparing just the state of the blockchain before a
transaction begins and after a transaction ends.
We are also able to inspect intermediate transition steps, but we are not
exploiting that feature yet.

\vspace{0.5em}
\myparagraph{Scilla.}
Scilla is a smart-contract language embedded in
Coq~\cite{Sergey.2018.Scilla} that allows some temporal reasoning
(see~\cite{Sergey.2018.TemporalProps}).
Scilla is an embedded domain-specific language in Coq which also abstracts smart
contracts as functions returning a list of operations.
The main difference between Multi and Scilla is that we do not present
a language to write smart-contract but use Coq functions directly.
We share the point where the effects of executing smart-contracts are
simple a list of operations that are propagated by the executer.
As Concert, we have a clean separation between the language of
smart-contracts and the machinery required to execute smart-contracts.
However, in our case, we decoupled the scheduler from the execution of
single instructions, and thus, we can implement different scheduling
strategies independently of the set of operations.

\vspace{0.5em}
\myparagraph{VerX.}
VerX is an automatic software verification tool that checks custom
functional properties of smart-contract entrypoints.
VerX works on a similar level to Mi-Cho-Coq, in the sense that they
prove functional properties of smart-contracts, but it is built to
be completely automatic and also to handle some multi-contract
interactions.
The interaction between smart-contracts comes from performing
analysis on the possible onchain behaviours of a set of
smart-contracts.
VerX restricts the analysis to a set of smart-contracts, \(S\), that
have a condition called \emph{effectively external callback free
  contracts}, which states that any behaviour generated by an
interaction between smart-contracts in set \(S\) that has an
external call is equivalent to a one without external
calls~\cite{Permenev.2020.VerX}.
This follows the lines of~\cite{Grossman.2018.EffectivelyCallbackFree}.
Because of that restriction, they can  reason about smart-contract, proving
PastLTL specifications, but it also restricts them to work in a \textbf{close
universe}.

\vspace{0.5em}
\myparagraph{SmartPulse.}
SmartPulse~\cite{Stephens.2021.Smartpulse} is another automatic
verification tool for smart-contracts.
The main goal is to verify temporal properties including some simple
liveness properties.
This tool is similar to \emph{VerX} but it is focused on proving
liveness properties of a single contract in a closed universe.
They do not support multi-contract interaction.

\section{Model of Computations}
\label{sec:model}

\myparagraph{Blockchain Model.}
We ignore the internals of the infrastructure of blockchain
implementations (like cryptographic primitives, consensus algorithms
or mempools) and focus exclusively on the model of computation that
blockchains offer to external users.
The blockchain is then abstracted by a partial map from addresses to
smart-contracts.
Smart-contracts are programs with some structure:
\begin{itemize}
\item Storage: a segment of memory that can only be modified by the
  smart-contract.
\item Balance: an attribute of contracts that indicates the amount
  of cryptocurrency stored in the contract.
\item The program code: a well-formed program that represents the
  implementation of the smart-contract.
\end{itemize}

The state of a smart-contract is a proper value of its storage plus
the balance its stores.
The model of computation consists of the sequential execution of
transactions, each of which is started by the invocation of an
operation.
In the current version, we ignore how gas or fees are paid or how new
currency is created during the evolution of the blockchain to pay the
bakers.
Smart-contracts can be executed upon request from an external user
that initiates a transaction or by the invocation from a running
contract.
Upon invocation, the blockchain evaluates the result of executing the
smart-contracts program following a given semantics producing effects
on the blockchain (further invocations) and changes on the
smart-contracts' storage or they may fail.

\vspace{0.5em}
\myparagraph{Open Universe.}
We introduce now the concept of \emph{universe of computation}.
Once a smart-contract has been installed on a blockchain, every other
entity in the blockchain can interact with it.
The smart-contract itself can invoke or be invoked by older or newer
contracts.
The case of smart-contracts invoking just older and well known
contracts can be useful sometimes but in general smart-contracts
may not know a priori who they are going to interact with.
This differs from conventional software where components are built
from well-known trustable components and the surface of interaction
with potentially malicious usage is small and well defined.
The classical way of programming exposes the internals of complex
software and leaves open attack vectors.
For example, to guarantee certain behaviour high-level
smart-contracts invoke low-level smart-contracts following a
protocol to logically guarantee a result.
However, malicious software \textbf{may not} follow such protocols possibly
breaking or leaving low-level smart-contracts in an incorrect state.
This open universe model of computation forces smart-contracts to
implement defensive mechanisms to prevent undesired
executions.

Most verification techniques and frameworks mentioned previously do
not take into care such assumption.
They operate under the idea that smart-contracts behave the way they
are supposed to, in the sense, that either they avoid external call
invocations by removing interactions or by assuming they are
interacting with good smart-contracts.
However, this is not the case, the blockchain is an aggressive
environment, a so called \emph{dark forest}~\cite{Robinson.2020.BlackForest}.
In this paper, we study this problem attempting to formalize properties
of smart-contracts operating under a more realistic (and
pessimistic) view of the world and also to develop new mechanisms or
features to explicitly guarantee that we are working under a safe
environment.
Such mechanisms could be implemented inside smart-contracts, but not every
mechanism can be implemented using current blockchain technologies, like
transaction monitors~\cite{Capretto.2022.TransactionMonitors,EIP.FlashLoan}.
%


\section{Formalization}
In this section, we describe the building blocks of our Coq library
implementation that allows us to reason about different blockchain
execution mechanisms.
Our goal is to study how smart-contracts interact with other
smart-contracts, and thus, we abstract away the internal execution of
the instructions of the smart-contract.
Moreover, we need a framework flexible enough to implement new features (i.e.
different execution models, scheduling strategies, etc) and, additionally, a
formal system to prove and verify properties of interactions between
smart-contracts implementing and using such features.
In short, we implemented a \emph{formal playground} simulating the
model of computation of blockchains.

We abstracted blockchains following the model described in
Section~\ref{sec:model} in the proof-assistant Coq.
We interpret smart-contracts as pure functions in the host language
Coq and every additional feature is implemented on top of
pure functions.

Smart-contracts are implemented as a structure with three
fields~(Listing~\ref{lst:smartcontract}): a storage, a balance, and a pure
function implementing the smart-contracts code.
\begin{listing}[!ht]
\begin{minted}[escapeinside=||]{coq}
Structure SmartContract (Ctx Param Storage Error Result : Type) : Type :=
  mkSmartContract {
      (* Storage *) _Sst : Storage ;
      (* Balance *) _Sbalance : |\(\mathbb{N}\)| ;
      (* Computation that result in an element of type |Result| *)
      _Sbody : Ctx |\(\to\)| Param |\(\to\)| Storage |\(\to\)| Error + (Result * Storage)
    }.
\end{minted}
\vspace{-1em}
\caption{Smart contract Definition}
\label{lst:smartcontract}
\end{listing}

Note that structure \solin{SmartContract} is highly parametric:
\begin{compactitem}
  \item Parameter \mintinline{coq}{Ctx} represents what smart-contracts can
observe about the blockchain and the execution model as: current block level, the
total balance of the transaction, who the sender and source are, etc.
\item Parameter \mintinline{coq}{Param} represents the parameters the
  body of the smart-contract expects to receive; using
  \mintinline{coq}{Param} we model the different  entrypoints of a
  contract.  
\item Parameter \mintinline{coq}{Storage} represents the storage of
  the smart-contract.
  \item Parameter \mintinline{coq}{Error} represents the type of errors that can
result from the execution of the smart-contract.
\item Parameter \mintinline{coq}{Result} represents the resulting type of smart
contracts, which in the Tezos model is a list of further operations.
\end{compactitem}
The type \solin{SmartContract} represents the most basic structure of
a smart-contract.
It is simply a structure with some storage, balance and a body.

The Smart-contracts body is modeled as a pure functions from the
current state of the blockchain and its storage to a sequence of
operations.
In this way, we abstract away concrete blockchain programming
languages or implementations.
Even though our formalization is based on the semantics of method
invocations in the Tezos blockchain, different programming language
can be modeled in this paradigm using standard compiler techniques
(essentially dividing a complex function with effects into its basic
blocks that are pure functions as modeled here).

\subsection{Execution}
The execution of a smart-contracts, aside from changes in the storage,
also produces a sequence of operations to be executed.
Therefore, we have to take care of two things: how to execute these
operations, and how to order the execution.
We split the execution model into two main pieces: a scheduler and an
executor.

\vspace{0.5em}
\myparagraph{Scheduler.}
The scheduler is in charge of the order of execution, adding new
operations the pending queue (either at the beginning or the end,
etc).
The scheduler is also in charge of creating new contexts.
Finally, it is in charge of building the graph/tree of transactions,
every information that descendants of an operation may share is kept
and organized by the scheduler.

\vspace{0.5em}
\myparagraph{Executer.}
The executer is in charge of executing an operation in a given
context, and it is the same independently of the evaluation order.
The most basic operation of an executor is smart-contract invocation,
which requires that the executor collects and builds the environment
in which such invocation should be executed.
The context is the blockchain state from the point of view of the
contract execution.
Another operation is smart-contract creation, which in this case it is going
to generate a modification to the blockchain, and communicate it to the
scheduler.

\vspace{0.5em}
\myparagraph{Operations.}
We assume the blockchain has a simple set of operations.
We start from a minimal set of operations that is simple enough to
enable smart-contracts interaction, and later add new operations as
needed afterward.

We begin our implementation with two operations:
\mintinline{coq}{Transfer} and \mintinline{coq}{Create_Contract}.
\begin{compactitem}
\item Operation \mintinline{coq}{Transfer} performs an invocation to a given
address while also sending money.
\item Operation \mintinline{coq}{Create_Contract} installs a new smart-contract
at an indicated address with an initial amount of balance and storage.
\end{compactitem}
%
%
\begin{minted}[escapeinside=!!]{coq}
Inductive EnvOps : Type :=
| Transfer : forall (T : Mich_Type),
    (* Parameter *) (Type_Interpret T) !\(\to\)!
    (* Amount to transfer *) Mutez !\sto!
    (* Contract address to invoke *) (Type_Interpret (ContractT T)) !\sto!
    EnvOps
| Create_Contract : forall (PTy StTy : Mich_Type),
    (* Pre-computed Address *) Address !\sto!
    (* Initial amount *) Mutez !\sto!
    (* Initial Storage *) (Type_Interpret StTy) !\sto!
    (* Body *) MichBodyTy PTy StTy (list EnvOps) !\sto!
    EnvOps.
  \end{minted}
  %
Where \solin{Mich_Type} is an enumeration type of the different data structures
supported by the blockchain, i.e. natural numbers, strings, etc.
In our case, since we are working close to the implementation of the Tezos
blockchain, we implement most of its data structures, and we represent them as
an inductive type \mintinline{coq}{Mich_Type}.
Using the previous operations, we can define smart-contracts simply as
the following structure:
\begin{minted}{coq}
Structure MichContract : Type := mkMich {
    (* Contract parameter type *) _Param : Mich_Type ;
    (* Storage type *) _Storage : Mich_Type;
    (* Contract body*)
    _Soul : SmartContract
                (TzCtxt _Param)
                (Type_Interpret _Param)
                (Type_Interpret _Storage)
                OError
                WritingContext;
    }.
\end{minted}
Essentially, we capture smart-contracts as their body plus
information about their types.
Hiding away the type information forces us to implement a lot of type
matching clauses when it comes to the execution of smart-contracts.
However, it enables us to represent the state of the blockchain simply
as a (partial) map of addresses to smart-contract.
\begin{minted}[escapeinside=||]{coq}
Definition TezosEnvironment := string |\sto| option MichContract.
\end{minted}
Given an operation, the executer is in charge of building the required
information to execute.
In the case of an invocation to an address \mintinline{coq}{addr}, the
executer looks up the address \coqin{addr} into the current
environment to see if there is a smart-contract matching the expected
type at that address, and in that case, executes its body to obtain
either a new storage and further operations or a fail.
In the case of a smart-contract creation operation, the executer is in charge of
checking that the address is actually free and updating the environment adding
such smart-contract.
Finally, the executer is also in charge of checking that
smart-contracts have enough balance to perform transactions and update
the current environment with the new balances.

We can characterize our executer as follows:
\begin{minted}[escapeinside=||]{coq}
Definition ExecuterTy : Type :=
   (* Input context information *) (ctx : ExecutionContext)
|\sto| (* Operation to execute*) (o : EnvOps)
|\sto| (* Current state *) (env : BCEnvironment)
|\sto| MFail (* possibly returning: *)
        (option(
          (* Address emitting new operations, next sender *) Address *
          (* Effects generated (new operations) *) WritingContext)
          * (* Updates to the environment *)
          (list (Address * MichContract))).
\end{minted}
Different executers exercising type \coqin{ExecuterTy} can interpret
operations in different ways.
Executers receive two arguments, \coqin{ctx} and \coqin{env},
representing the execution context and the environment of the
blockchain, respectively, and in return, provide the modifications to
the environment and possibly a list of new operations.
Note that \coqin{ExecuterTy} leaves some proofs obligations if we want to
simulate current blockchains, i.e. we need to show that \coqin{ExecuterTy} does
not modifies or upgrades exiting smart-contracts' code~(see
Section~\ref{sec:correctness}).

Schedulers are in charge of gluing
together the effects generated by the execution of operations in the current
blockchain.
We model them in Coq as a type listed in Listing~\ref{lst:schedulers} where
\coqin{SchedulingStrategy} implements the execution order to follow.
In other words, schedulers keep track of the evolution of the
state of the blockchain while managing the pending queue of
operations.
Schedulers take the first operation on the pending queue, build
the information required by the executor, and pass everything to an
executer.
When executers return, schedulers take the resulting operations and
updates to the current state of the blockchain.

\begin{listing}[!ht]
\begin{minted}[escapeinside=||]{coq}
Definition Scheduler : Type
  := (* Strategy *) SchedulingStrategy
  |\sto| (* External user *) Address
  |\sto| (* Executor *) ExecuterTy
  |\sto| (* Current environment *) BCEnvironment
  |\sto| (* Time *) Timestamp
  |\sto| (* Pending Execution list *) list (list EnvOps * ExecutionContext)
  |\sto| (MFail BCEnvironment * Timestamp).
\end{minted}
\vspace{-1em}
\caption{Schedulers type}
\label{lst:schedulers}
\end{listing}

The computation of a transaction begins with an external user (outside the
blockchain) posting one or more operations to be executed, defined in
Listing~\ref{lst:signedtrans}.
\begin{listing}[!ht]
\begin{minted}{coq}
Structure SignedTrans : Type := mkSignedTrans {
      _author : Address; _trans : list EnvOps
    }.
\end{minted}
\vspace{-1em}
\caption{Signed transactions definition.}
\label{lst:signedtrans}
\end{listing}
The initial transaction is given to the scheduler, which also receives
a scheduler strategy, an executer, and a context to compute the
transaction and its descendants operations.
The result is a pair composed of a possible new environment and the
next timestamp.
We need timestamps to represent the passage of time, and thus, time
progresses even in the case that a transaction is reverted.
In practice, the scheduler strategy is fixed for a given blockchain.

Since blocks in the blockchain are just sequences of signed
transactions, \coqin{SignedTrans}, we can generate arbitrary traces
with systems like QuickChick~\cite{Paraskevopoulou.2015.QuickChick}.
Given a logical program (reflected in a set of smart-contracts), we can codify
the possible logical operations in an inductive type in Coq.
Therefore, we can generate a sequence of actions translating the logical steps
into transactions in the blockchain and verify that the smart-contracts do not
reach an invalid state.


\subsection{Proof of Correctness}\label{sec:correctness}
%
%
We can define a specification of how a proper blockchain should behave and check
that our implementation follows the specification.
For example, a basic property is \emph{no-double spending} which
states that transfers (remote contract invocations) are paid once,
i.e. the sender is not charged twice for the same operation.
We can go even further and prove that executing a transfer does exactly what it
is supposed to do~(Listing~\ref{lst:transferok}), i.e. invokes another
smart-contract, executes its code, deduce the expected amount from the sender's
account, and adds it to the destination's account, or fail (in which case the
transfer has no effect).

\begin{listing}[!ht]
\begin{minted}[escapeinside=||]{coq}
Lemma SimpleTransferCheck :
  forall callerContract calleeContract parameterTy BCCtxt send
    (parameter : Type_Interpret parameterTy)
    (storage storage' : Type_Interpret (_Storage calleeContract))
    (contractContext : TzCtxt parameterTy),
    successWith (ops, (caller', callee'))
                (SimpleTransfer callerContract send calleeContract)
    |\sto| _st calleeContract |\sequiv| storage
    |\sto| successWith (ops, storage') (exec calleContract BCCtxt parameter)
    |\sand| ((_balance callerContract) - send) |\sequiv| _balance caller'
    |\sand| ((_balance calleeContract) + send) |\sequiv| _balance callee'
    |\sand|  _st callee' |\sequiv| storage'.
\end{minted}
\vspace{-1em}
\caption{Transfer is correct.}
\label{lst:transferok}
\end{listing}

An alternative approach would be to define a small step inductive
relation defining how blockchains should behave and prove that the
scheduler follows it step by step.
The framework Concert~\cite{Annenkov.2020.ConCert} follows that
approach.

\subsection{Multi-Contract Interaction Proofs}
The most important part of our framework is that we can simulate
executions of smart-contracts and inspect the effects generated by
smart-contract interactions.
In other words, we have a big-step semantics of blockchain operations
where we can study how smart-contracts using different mechanisms
(i.e. BFS/DFS, etc) interact with each other.
We can build proofs either by observing the evolution of the
transaction execution operation by operation, or analyzing its final
state after the transaction terminates.
In other words, we have a definition of observational equivalence of smart
contracts modulo the particular blockchain employed as evaluator.

This is extremely useful because we can abstract away entire
smart-contracts and event simulate the more realistic scenario: a
demonic environment.
Either we know the code of smart-contract and we can predicate over
these code during the proof, or we do not have these code, which
requires reasoning with universal quantification over all possible
smart-contracts.
In other words, to prove that smart-contracts are prepared to operate
properly in the open universe of the blockchain requires to reason
about the interactions with all possible contracts.

We can model angelic computations by expanding our known universe of
smart-contracts simply by implementing smart-contract on our framework
and having them installed in the blockchain inside a simulation.


\section{Conclusion}

In this paper, we present Multi, a formal playground to reason about smart
multi-contract interaction and to study features of the blockchain before
deployment.
Additional features and mechanisms are described in
Appendix~\ref{app:restrictions} and Appendix~\ref{app:bundles}
where we introduce the idea of \emph{Bundles} of operations: semantic
restrictions on the execution of a sequence of operations.
Our framework, based on the Tezos blockchain, is very general and
allows us to reason about different execution orders, abstracting away
each operation on a contract by a pure function whose output is either
a failure or the changes in the local storage plus further operations.

Future work includes:
\begin{compactitem}
  \item Examples and study cases: implement and study complex use cases.
  \item Integrate Multi to the Tezos formal ecosystem and study 
interactions with Concert and Mi-Cho-Coq.
  \item Implement additional features, e.g. transaction monitors, views, etc,
and study how they interact between each other.
  \item Design and implement a DSL to easily encode specific smart-contracts
easing the translation from existing languages into Coq functions.
\item Write more expressive smart contract types following the steps of Concert since
Coq functions are more general than the contracts accepted by most blockchains (like Tezos).
\item Implement complex features as \emph{tickets}/NFT using some mechanisms
(like monads) to better capture the space of functions that represent
smart-contracts.
  \end{compactitem}

Finally, we aspire to implement a richer specification language using
ATL~\cite{Alur.2002.ATL} to describe the interaction between smart-contracts and
fully verify their specification in Coq.
The idea consists in describing programs as interactions between agents (i.e.
smart-contracts) where agents cooperatively guarantee certain properties or
exercise certain rights.
At the semantic level, we would connect the evaluation of smart-contracts in a
blockchain with their semantic given by ATL and concurrent games.
In other words, with Multi, we can interact between a rich specification
language of smart-contracts and their behaviour defined by the execution of
blokchains.

%
%


\bibliographystyle{plainurl}
\bibliography{bibfile}

\appendix

\section{Angelic/Demonic}\label{sec:ang:dem}

Given the open universe nature of blockchains, smart-contracts are forced to
identify who are they interacting with.
Programmers when they are designing complex software do not think that they are
in a dangerous and aggressive environment, as it is now, and simply think that
smart-contracts will interact with good pieces of software doing what they are
supposed to do.
However, as we saw before, this may not be true.

In this section, we present a new characterization when it comes to classifying
the interaction between multiple smart-contracts.
We call this characterization \emph{Angelic/Demonic} where we mark
smart-contracts as \emph{angelic} when they do what they are supposed to do, or
as \emph{demonic} when we cannot assume anything about their behaviour, and
thus, we cannot predict nor predicate about their behaviour.
Note that this is not a property enforced by blockchains, but it is more of a
mindset at the moment of designing complex software that is going to run on the
blockchain.

There are essentially two basic models to reason about multi-contract interaction:
\begin{description}
    \item[Closed World Assumption:]\label{close:world} every smart-contract knows and trusts the
smart-contracts that it is invoking (directly and transitively).
        In particular, every smart-contract C only invokes contracts that are
older than C and whose properties are known.
    \item[Open World Assumption:]\label{open:world} every contract C runs in an
adversarial environment and smart-contracts should protect against possible
\emph{evil} smart-contracts.
\end{description}

A closed world assumption is feasible on many occasions because of the public
and \emph{immutable}\footnote{Although it is possible to implement mutable and
upgradable smart-contracts, this is not the general case, and even if the nature of the smart-contract was to mutate this would be known by the
invoker.} character of the blockchain.
Since everything is public and smart-contracts do not change, as smart-contract
developers, we can observe the state and code of smart-contracts that we are
going to interact with and decide if they are \emph{angelic}, i.e. if they do
what they are supposed to do.

Note that ``the angelic state'' is fragile and it may change.
For example, assume we invoke a smart-contract \(B\) that in turn invokes
another smart-contract whose address \(addr\) is stored in \(B\)'s storage.
As we are about to submit our smart-contracts to the blockchain, we can explore
and decide that \(B\) and the current \(addr\) are angelic.
However, eventually, \(B\) may change it to another smart-contract \(addr'\)
that may also be angelic to \(B\), or \(B\) is protected towards possible
attacks from \(addr'\), but it may open an attack on our smart-contract.

The second option, an open world assumption, is a more real situation and
sometimes the only possible case for certain smart-contracts.
One of the most prominent cases is exchange houses: let \emph{Dex} be a
smart-contract that is always willing to exchange token \(A\) for token \(B\)
for a certain fee in behave of a set of investors.
In this case, the smart-contract \emph{Dex} is doomed to interact with unknown
addresses.

Another example is that we can implement a call-and-return model using
\emph{continuation passing style} between smart-contracts in BFS blockchains.
However, implementing such interactions between smart-contracts requires to
assume that \emph{every smart-contracts is going to behave accordingly}, and
thus, we are under an angelic assumption.
Therefore, we need a framework that can handle angelic and demonic assumptions.


\section{Bundles of Operations}\label{app:bundles}

In this section, we introduce the concept of \emph{bundles of operations} high
level restrictions on how we want a sequence of operations to be executed.
For example, we can abstract away what is important about a scheduler following
a BFS strategy: atomicity of a sequence of operations.
In other words, the operations generated by a smart-contract are going to be
executed one after another without other smart-contracts injecting operations
between them.
%
%

A \emph{bundle} is a semantic condition (or restriction) on the execution of a
sequence of operations.
Instead of forcing \emph{the whole blockchain} to use a particular execution
order, we theorize on having a domain-specific language (DSL) describing
how we would want to execute a set of operations.
In other words, we would like to predicate on how operations are to be
executed explicitly, either by assuming a BFS/DFS or other mechanisms.

\subsection{Atomic Sequence}
Given a sequence of operations \(\langle s_{0}, s_{1}, \ldots, s_{n}\rangle\), we want them to be
executed atomically without interleaving operations independently of the
execution order followed by the scheduler.
BFS schedulers respect such bundle by definition, while DFS schedulers should
check that the effects generated by each \(s_{i}\) with \(i \leq n\) does not
affect the rest of the smart-contracts.

\subsection{Contexts}
The call and return pattern enables us to reason about units of functionality,
in the sense, that when we invoke a method in a smart-contract is because we
expect a result independently of how many other functions that method is
invoking.
When we program smart-contracts under the demonic assumption, where giving
control to other (possibly unknown) smart-contract may result in an attack, we
want to encapsulate their behaviour while still interacting with them to obtain
some functionality.

Independently of the execution order, we can devise an encapsulation mechanism
enabling us to reason about the functionality of external invocations in a
\emph{context}.
The general idea is to encapsulate the execution of smart-contracts and all of
its descendant operations in a \emph{context}.
Instead of having a pending queue of operations, we would have a sequence of
pending queues, each one representing an encapsulated context.
Operationally, each context is completely executed before passing to the next.
Contexts give us the ability to invoke functions and execute them as if they
were the only procedures being executed in the machine, i.e. in a completely
isolated context.

\begin{example}
  Let \(A\) and \(B\) be two smart-contracts such that
  the result of executing \(A\) is two operations \([A_{1},A_{2}]\),
  while the result of executing \(B\) is just \([B_{1}]\).
  Moreover, operations \(A_{1},A_{2}\) do not generate new operations.

  Assuming we have a pending queue formed by a context invocation to \(A\)
followed by a normal invocation to \(B\), we will have the following execution
sequence:
\[
  [ [A] , B ] \leadsto [ [A_{1},A_{2}] , B ] \leadsto [ [A_{2}] , B ] \leadsto [[] , B] \equiv [B] \leadsto \ldots
\]
  \end{example}


Implementing contexts is easy and very useful to encapsulate functionality.
However, this brings some questions: how are contexts created?  who
creates them?
From the point of view of defense programming, we have two possible answers:
\begin{description}
    \item[Caller contextual call:] upon invoking a remote procedure, the caller
can specify the execution to be encapsulated in a context.
This mechanism protects the callee since the new procedure cannot inject
operations interleaving the ones already on the pending queue (as a DFS
blockchain would do).
    \item[Callee contextual call:] when invoked, the callee internally decides
if its functions are to be executed in a context.
This mechanism enables the function being called to assume that the pending
execution queue is empty and nothing is going to modify it aside from itself or
the invoked smart-contracts.
  \end{description}


\section{Restricting Smart-Contracts Interaction}\label{app:restrictions}

We implemented two kinds of restrictions: one where the blockchain enters into a
mode where the smart-contract interactions are not allowed, and another where we
can reduce the set of addresses that can be invoked.

\vspace{0.5em}
\myparagraph{End of Interactions.}
The executor only accepts transactions from and to the same smart-contract.

\vspace{0.5em}
\myparagraph{Address Universe.}
We can dynamically restrict the universe of addresses that smart-contracts (and
their descendants) can invoke, either by restricting the known universe of
addresses or by specifying addresses that cannot be invoked.
In other words, we would have two sets of addresses:
\begin{description}
    \item[Allow addresses:] the set of addresses that can be invoked during execution.
        Invoking an address outside this set will force the transaction to fail.
    \item[Block addresses:] the set of addresses that cannot be invoked during execution.
        Invoking one of these addresses will force the transaction to fail.
  \end{description}

Both mechanisms suggest the addition of a shared state between a smart-contract
and its descendants during the execution of smart-contracts.
If we see transaction executions as trees, we can add restrictions to such
tree.
Moreover, we can analyze \emph{transaction trees} to restrict or predict the
behaviour of smart-contracts.


\end{document}